\documentclass[12pt]{article}
\usepackage{epsfig}
\usepackage{amsmath}  
\usepackage{hhline}
\usepackage{amssymb}
\usepackage{times}
\usepackage{cite}
\usepackage{psfrag}  
\usepackage{multirow}
\usepackage{rotating}

\newlength{\dinwidth}
\newlength{\dinmargin}
\begin{document}  

\newcommand{\thstarhad}{$\theta^{q}_{W^{*}}$~} 
\newcommand{\ktug}{$\kappa_{tu\gamma}$~}
\newcommand{\vtuz}{$V_{tuZ}$~}

\hyphenation{experiment} 
 
\def\Journal#1#2#3#4{{#1} {\bf #2} (#3) #4}
\def\NCA{\em Nuovo Cimento}
\def\NIM{\em Nucl. Instrum. Methods}
\def\NIMA{{\em Nucl. Instrum. Methods} {\bf A}}
\def\NPB{{\em Nucl. Phys.}   {\bf B}}
\def\PLB{{\em Phys. Lett.}   {\bf B}}
\def\PRL{\em Phys. Rev. Lett.}
\def\PRD{{\em Phys. Rev.}    {\bf D}}
\def\ZPC{{\em Z. Phys.}      {\bf C}}
\def\EJC{{\em Eur. Phys. J.} {\bf C}}
\def\CPC{\em Comp. Phys. Commun.}


\begin{titlepage}

\noindent
\begin{flushleft}
{\tt DESY 09-050    \hfill    ISSN 0418-9833} \\
{\tt June 2009}                  \\
\end{flushleft}



\vspace{2cm}
\begin{center}
\begin{Large}

{\bf Search for Single Top Quark Production at HERA}

\vspace{2cm}

H1 Collaboration

\end{Large}
\end{center}

\vspace{2cm}

\begin{abstract}
\noindent

A search for single top quark production is performed in the full $e^\pm p$ 
data sample collected by the H1 experiment at HERA,
corresponding to an integrated luminosity of $474$~pb$^{-1}$. 
Decays of top quarks into a\, $b$~quark 
and a $W$~boson with subsequent leptonic or hadronic decay 
of the $W$ are investigated. 
A multivariate analysis is performed to 
discriminate top quark production from Standard Model background processes.
An upper limit on the top quark production cross section 
via flavour changing neutral current processes
$\sigma (ep\rightarrow etX) < 0.25$~pb is established at $95\%$~CL. 
Limits on the anomalous 
coupling $\kappa_{tu\gamma}$ 
are derived. 

\end{abstract}

\vspace{1.5cm}

\begin{center}
Accepted by \PLB
\end{center}

\end{titlepage}

%
%
%
\begin{flushleft}

F.D.~Aaron$^{5,49}$,           
M.~Aldaya~Martin$^{11}$,       
C.~Alexa$^{5}$,                
K.~Alimujiang$^{11}$,          
V.~Andreev$^{25}$,             
B.~Antunovic$^{11}$,           
A.~Asmone$^{33}$,              
S.~Backovic$^{30}$,            
A.~Baghdasaryan$^{38}$,        
E.~Barrelet$^{29}$
% ...

\bigskip{\it
 $ ^{1}$ I. Physikalisches Institut der RWTH, Aachen, Germany$^{ a}$ \\
% ...
$ ^{5}$ National Institute for Physics and Nuclear Engineering (NIPNE) ,
          Bucharest, Romania \\
% ...
$ ^{11}$ DESY, Hamburg, Germany \\
% ...
$ ^{25}$ Lebedev Physical Institute, Moscow, Russia$^{ e}$ \\
% ...
$ ^{29}$ LPNHE, Universit\'{e}s Paris VI and VII, IN2P3-CNRS,
          Paris, France \\
 $ ^{30}$ Faculty of Science, University of Montenegro,
          Podgorica, Montenegro$^{ e}$ \\
% ...
$ ^{33}$ Dipartimento di Fisica Universit\`a di Roma Tre
          and INFN Roma~3, Roma, Italy \\
% ...
$ ^{38}$ Yerevan Physics Institute, Yerevan, Armenia \\
% ...
$ ^{49}$ Also at Faculty of Physics, University of Bucharest,
          Bucharest, Romania \\
% ...
$ ^a$ Supported by the Bundesministerium f\"ur Bildung und Forschung, FRG,
      under contract numbers 05 H1 1GUA /1, 05 H1 1PAA /1, 05 H1 1PAB /9,
      05 H1 1PEA /6, 05 H1 1VHA /7 and 05 H1 1VHB /5 \\
% ...
$ ^e$ Supported by the Deutsche Forschungsgemeinschaft \\
% ...
}
\end{flushleft}
%

\newpage

\section{Introduction}
\label{ch:introduction}

%
Top quarks are of particular interest in searches for new physics
because 
their mass is close to the electroweak scale. 
In $e^{\pm}p$ collisions at HERA the production of single 
top quarks is kinematically possible due to the large centre of mass 
energy of up to $\sqrt s=319$~GeV.
Within the Standard Model (SM) the production of top quarks at HERA
is however strongly suppressed.
Therefore the observation of single top quark production would be a clear indication
of new physics.
In several extensions of the SM the top quark 
is predicted to undergo flavour changing neutral 
current (FCNC) interactions, which may lead to a sizeable 
top quark production cross section at HERA~\cite{Fritzsch:1999rd,AguilarSaavedra:2004wm}.

A search for single top quark production is performed using the full $e^\pm p$
data sample collected by the H1 experiment at HERA.
The data correspond to an integrated luminosity of $474$~pb$^{-1}$,
including $36$~pb$^{-1}$ of data taken at $\sqrt s=301$~GeV.
The analysis is inspired by and supersedes the previous H1 search for single 
top quark production 
based on an integrated luminosity of $118$~pb$^{-1}$~\cite{Aktas:2003yd}. 
Anomalous single top quark production also has been
investigated by the ZEUS collaboration \cite{Chekanov:2003yt} and by
the LEP experiments \cite{Achard:2002vv}.

In the present analysis, single top quark production is detected via the decay of the top quark
$t \rightarrow b W$.
In the case of leptonic decays of the $W$ boson,  $W \rightarrow \ell \nu$, 
the signature is a charged lepton (electron or muon) and missing transverse momentum,
accompanied by 
a hadronic final state $X$ with
a high transverse momentum ($P_T$) due to the $b$ quark. 
Events of this topology have been studied in a recently published 
analysis of events with isolated 
leptons and missing transverse momentum by H1~\cite{IsoLep:2009}. 
In the case of the hadronic $W$ boson decay channel,
$W \rightarrow q \bar{q} $, 
the signature of single top quark production consists of 
three high $P_T$ jets with an invariant mass 
compatible with the top quark mass.
A multivariate discriminant based on a neural network is used
to differentiate top quark production from SM background. 
A possible contribution from a top quark signal is extracted using 
the method of fractional event counting~\cite{Bock:2004xz}.

\section{Single Top Quark Production}

\begin{figure}
  \begin{center}      
      \includegraphics[width=0.6\textwidth]{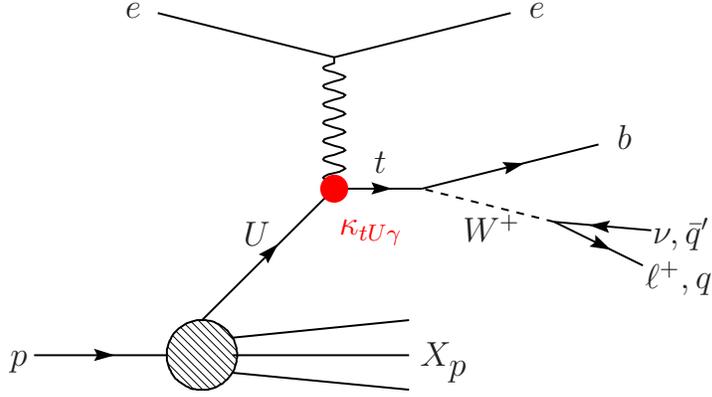}
  \end{center}   
  \caption{ Anomalous single top quark production in $ep$ collisions
  via a FCNC coupling $\kappa_{tu\gamma}$.
  }
  \label{fig:anotop_fd}
\end{figure}

The dominant process of SM single top quark production at HERA is via charged current (CC)
production $ep \rightarrow \nu t b X$. 
The cross section for this process has been
estimated~\cite{Stelzer:1998ni} 
to be $\mathcal{O}(1)$~fb and is thus not observable with the available
integrated luminosity.
Anomalous single top quark production within FCNC models, 
where the coupling
of a top quark with an up-type quark $U$ via a photon is described by a
coupling $\kappa_{tU\gamma}$, is illustrated in figure~\ref{fig:anotop_fd}.
This process, as well as FCNC top quark interactions involving
vector couplings to a $Z$ boson $V_{tUZ}$,
can be described by an effective Lagrangian~\cite{Han:1998yr}:
\begin{equation}
\label{eq:lagrangian}
 {\cal{L}}_{eff}^{FCNC} =  \sum_{U = u,c} 
  \frac{e e_U}{2\Lambda}  \kappa_{tU\gamma} \bar{t} \sigma_{\mu \nu} A^{\mu\nu} U \nonumber 
 + \frac{g}{2 \cos \theta_W} V_{tUZ} \bar{t} \gamma_{\mu} U Z^{\mu}
 \quad + \quad {\mbox {h.c.}} 
\ ,
\end{equation} 
where $\sigma_{\mu \nu} = (i/2) \left[ \gamma^{\mu}, \gamma^{\nu} \right]$,
$\theta_W$ is the Weinberg angle, 
$e$ and $g$ are the couplings to the electromagnetic and weak gauge groups 
with $U(1)$ and $SU(2)$ symmetries, respectively, 
$e_U$ is the electric charge of up-type quarks, 
$A^{\mu\nu}$ is the field strength tensor of the photon,
$Z^{\mu}$ is the vector field of the $Z$ boson
and $\Lambda$ is a scale parameter.

As the top quark mass is comparable to the  $ep$ centre of mass energy at HERA,
the interacting parton in the proton must be at high Bj{\o}rken-$x$.
Contributions from the charm quark are therefore neglected 
($\kappa_{tc\gamma} \equiv V_{tcZ} \equiv 0$) 
since the $c$ quark density in the proton is low at high Bj{\o}rken-$x$.
Similarly, the production of anti-top quarks is neglected,
as this process involves sea anti-quarks in the initial state.
%

%
The simulation of an anomalous single top quark signal is done using the
event generator ANOTOP~\cite{Aktas:2003yd}, which uses the leading 
order (LO) matrix elements of the complete 
$e + q \rightarrow e + t \rightarrow e + b + W \rightarrow  e + b + f + \bar{f'}$
process as obtained from the CompHEP program~\cite{COMPHEP}.
%
%
ANOTOP is used to calculate the production cross section and
to study decays of top quarks in the H1 detector.
Only top quark decays $t \rightarrow b W$ are considered
%
as suggested by the strict limits on other possible top quark
decays~\cite{Aaltonen:2008aaa,Abe:1997fz}.

The anomalous single top quark production cross section can then 
be parametrised as:
\begin{equation}
\label{anotop_xsec}
\sigma ( ep \rightarrow etX, \sqrt{s}) 
= c_{\gamma} \cdot \kappa_{tu\gamma}^2 + c_{Z} \cdot V_{tuZ}^2
	+ c_{\gamma Z} \cdot \kappa_{tu\gamma} \cdot V_{tuZ}.
\end{equation}
The coefficients $c_{\gamma}$ and $c_{Z}$
are determined with ANOTOP
at $\sqrt{s} = 319$~GeV.
In this analysis $m_{\rm{top}}$ is set to $175$~GeV and 
by convention the scale parameter $\Lambda$ is fixed to $m_{\rm{top}}$.
If $\Lambda \equiv m_{\rm{top}}$ is reduced to $170$~GeV,
the cross section increases by $25\%$, mainly due to an increased phase space
for the production of top quarks.
This mass range encompasses the current top quark mass determination
of CDF and D{\O} 
$m_{\rm{top}} = 172.4 \pm 1.2$~GeV~\cite{:2008vn}
and corresponds to a coefficient range from $c_{\gamma}= 7.53$~pb to $9.41$~pb 
and $c_{Z} = 0.26$~pb to $0.32$~pb.
The values for $c_{\gamma}$ include next to leading order (NLO) 
corrections~\cite{Belyaev:2001hf},
which increase the LO cross section by $17\%$.
Including these NLO corrections results in an uncertainty related to the choice of the
renormalisation and factorisation scales of about $5\%$.
Since the contribution of the $Z$ boson is small and no competitive sensitivity is expected,
$Z$-exchange is neglected in this analysis and only the $\kappa_{tu\gamma}$ coupling is 
considered.
The interference term with coefficient $c_{\gamma Z}$ contributes less than $1\%$ of 
the total cross section and is therefore also neglected. 
 

\section{Standard Model Background Processes}
\label{ch:sm}

The contributions from SM processes to the background in the leptonic decay channels
were studied extensively in the analysis of events with isolated leptons and
missing transverse momentum~\cite{IsoLep:2009}
and are only briefly described here.
The main SM background process is the 
production of single $W$ bosons.
Further SM background processes may mimic the top quark signature through
misidentification or mismeasurement.
In the hadronic decay channel the SM background prediction is
dominated by multi-jet events produced in photoproduction or
neutral current (NC) deep-inelastic scattering (DIS).

Production of single $W$ bosons is
modelled by the EPVEC generator~\cite{Baur:1991pp}. 
NLO QCD corrections are taken into account by reweighting the
generated events~\cite{Diener:2002if,Diener:2003df}.
CC DIS events are simulated using the
DJANGO~\cite{Schuler:yg} generator, which includes first order
leptonic QED radiative corrections based on HERACLES~\cite{Kwiatkowski:1990es}.
The production of two or more jets in DJANGO is accounted for using
the colour-dipole~model~\cite{Lonnblad:1992tz}.
The RAPGAP~\cite{Jung:1993gf} event generator, which implements the 
Born, QCD Compton and Boson Gluon Fusion matrix elements, 
is used to model NC DIS events.
Direct and resolved photoproduction of jets and prompt photon
production are simulated using the PYTHIA~\cite{Sjostrand:2000wi}
event generator.
The simulation is based on Born level hard scattering matrix elements
with radiative QED corrections.
In RAPGAP and PYTHIA, jet production from higher order QCD radiation
is simulated using leading logarithmic parton showers while
hadronisation is modelled with Lund string fragmentation.
The leading order MC prediction from NC DIS and photoproduction
processes with two or more high transverse momentum jets is scaled by
a factor of $1.2$, to account for missing higher order QCD
contributions in the MC generators~\cite{Adloff:2002au,Collaboration:2009ww}.
Further small contributions to the SM background originate from
lepton pair production, simulated using the 
GRAPE generator~\cite{Abe:2000cv}, and from photon wide 
angle bremsstrahlung, simulated in the WABGEN framework~\cite{Berger:1998kp}.

Generated signal and background events are passed through the
GEANT~\cite{Brun:1987ma} based simulation of the H1 detector, which
takes into account the running conditions of the different data taking
periods, and are reconstructed and analysed using the same program
chain as for the data.

\section{Experimental Conditions}
\label{ch:experiment}

A detailed description of the H1 experiment can be found
in~\cite{Abt:1997}. 
Only the detector components relevant to the present analysis are briefly described here.
The origin of the H1 coordinate system is the nominal $ep$ interaction
point, with the direction of the proton beam defining the positive
$z$-axis (forward region). 
Transverse momentum is measured in the \mbox{$xy$-plane}.
The pseudorapidity $\eta$ is related to the polar angle
$\theta$ by $\eta = -\ln \, \tan (\theta/2)$.
Tracking is provided by the central \mbox{($20^\circ<\theta<160^\circ$)} and 
forward \mbox{($7^\circ<\theta<25^\circ$)} tracking detectors.
They are used to measure
charged particle trajectories, to reconstruct the interaction vertex
and to complement the measurement of hadronic energies.
The tracking detectors are surrounded by a finely segmented Liquid Argon (LAr) 
calorimeter~\cite{Andrieu:1993kh} 
that covers the
polar angle range $4^\circ < \theta < 154^\circ$. 
Electromagnetic shower energies are measured with a
precision of $\sigma (E)/E = 12\%/ \sqrt{E/\mbox{GeV}} \oplus 1\%$ and
hadronic energies with $\sigma (E)/E = 50\%/\sqrt{E/\mbox{GeV}} \oplus
2\%$, as measured in test beams~\cite{Andrieu:1993tz,Andrieu:1994yn}.
In the backward region, energy measurements are provided by a
lead/scintillating-fibre (SpaCal) calorimeter\footnote{This device
was installed in 1995, replacing a lead-scintillator ``sandwich`` calorimeter~\cite{Abt:1997}. }~\cite{Appuhn:1996na}
covering the range $155^\circ < \theta < 178^\circ$.
The LAr calorimeter and inner tracking detectors are enclosed in a
super-conducting magnetic coil with a field strength of $1.16$~T.
From the curvature of charged particle trajectories in the magnetic
field, the central tracking system provides transverse momentum 
measurements with a resolution of $\sigma_{P_T}/P_T = 0.005 P_T /
\rm{GeV} \oplus 0.015$~\cite{Kleinwort:2006zz}.
The return yoke of the magnetic coil is the outermost part of the
detector and is equipped with streamer tubes forming the central muon
detector ($4^\circ < \theta < 171^\circ$).
\index{\footnote{}}%
In the forward region of the detector ($3^\circ < \theta < 17^\circ$) 
a set of drift chambers detects muons and measures their momenta
using an iron toroidal magnet.
The luminosity is determined from the rate of the Bethe-Heitler
process $ep \rightarrow ep \gamma$, measured in a photon detector
located close to the beam pipe at $z=-103~{\rm m}$, in the backward
direction.


The LAr calorimeter provides the main trigger for events in this
analysis.
The trigger efficiency is almost 100\% for events with an electron
with an energy above $10$~GeV.
Events with muons are triggered by
an imbalance in transverse momentum measured in the calorimeter
$P_{T}^{\rm{calo}}$.
The trigger efficiency is~$\sim 60\%$ for
$P_{T}^{\rm{calo}}=12$~GeV, rising to~$\simeq 98\%$ for
$P_{T}^{\rm{calo}}>25$~GeV.
Events with jets 
have a trigger efficiency of nearly $100\%$ for events 
with $P_T^{\rm jet}>25$~GeV~\cite{matti}.


In order to remove background events induced by cosmic rays and
other non-$ep$ sources, the event vertex is required to be
reconstructed within $\pm 35$~cm in $z$ of the average interaction
point.
In addition, topological filters and timing vetoes are applied.

\section{Data Analysis}
\label{ch:reco}

\subsection{Particle Identification and Event Reconstruction}
\label{sec:kine}

Particle identification and hadronic final state reconstruction are
described in detail in~\cite{IsoLep:2009} and are only briefly summarised here.
Electrons and photons are characterised~\cite{pbruel} by compact and isolated 
electromagnetic clusters in the LAr calorimeter or SpaCal.
Muon identification
is based on a track in the
inner tracking systems associated with signals in the muon
detectors~\cite{Aktas:2003sz,Aaron:2008jh}.
Calorimeter energy deposits and tracks not previously identified as
electron, photon or muon candidates are used to form combined cluster-track
objects, from which the hadronic final state is
reconstructed~\cite{matti,benji}.
Jets are reconstructed from these combined cluster-track objects 
using an inclusive $k_T$ algorithm~\cite{Ellis:1993tq,Catani:1993hr} with a
minimum transverse momentum of $4$~GeV.
The missing transverse momentum $P^{\rm miss}_{T}$ in the event is 
derived from all detected particles and energy deposits in the event.

Strict isolation criteria are applied to electron and muon candidates.
For electrons a track with a distance of closest 
approach (DCA) to the cluster of less than $12$~cm is required
to reject photons.
The calorimetric energy measured within a distance in 
the pseudorapidity-azimuth $(\eta - \phi)$ plane 
$D=\sqrt{\Delta \eta^2 + \Delta \phi^2} < 0.5$
around the electron cluster is required to be below $3\%$ of its energy.
Furthermore, electrons are required to be isolated from
jets by a distance $D(\rm{e;jet}) > 1.0$ to any jet axis,
and in the region $\theta>45^{\circ}$
by a distance $D(\rm{e;track})>0.5$ from any track. 
A muon candidate may have no
more than $5$~GeV energy deposited in a cylinder, centred on the muon track
direction, of radius $25$~cm and $50$~cm in the electromagnetic and
hadronic sections of the LAr calorimeter, respectively.
Muon candidates are required to be separated from any jet by a distance
$D(\rm{\mu;jet})>1.0$ and from any track by $D(\rm{\mu;track})>0.5$.

\subsection{Systematic Uncertainties}
\label{sec:systematics}
%
The following experimental systematic uncertainties are considered:
 \begin{itemize}
 \item The uncertainty on the electromagnetic energy scale varies
   depending on the polar angle from $0.7\%$ 
   to $2\%$. The polar angle measurement uncertainty is $3$~mrad for
   electromagnetic clusters.
 \item The scale uncertainty on the transverse momentum of high $P_T$
   muons is $2.5\%$. The uncertainty on the muon polar angle measurement
   is $3$~mrad.
 \item The hadronic energy scale is known within $1.5\%$~\cite{maxime}.
 The uncertainty on the jet polar angle measurement is $10$~mrad~\cite{matti}.
 \item The uncertainty on the trigger efficiency is $3\%$.
 \item The luminosity measurement has an uncertainty of $3\%$.
 \end{itemize}
The effect of the above systematic uncertainties on the MC expectation
is determined by varying the experimental quantities by $\pm 1$
standard deviation in the MC samples and propagating these variations
through the whole analysis.

Additional model uncertainties are attributed to the SM MC
generators described in section~\ref{ch:sm}.
A theoretical uncertainty of $15\%$ 
is used for the predicted contributions from EPVEC~\cite{Diener:2002if}.
In the electron and muon channels the CC background contribution, 
which is modelled using DJANGO, is attributed a systematic error of 
$50\%$~\cite{IsoLep:2009}.
The contributions from
background processes modelled using RAPGAP, PYTHIA, GRAPE
and WABGEN are attributed $30\%$ model uncertainties~\cite{IsoLep:2009}.
In the hadronic channel, the background processes with multi-jet final states 
modelled using RAPGAP and PYTHIA are attributed a $10\%$ model uncertainty
determined from the comparison to data in an extended phase space.

\subsection{Electron and Muon Channels}
\label{sec:elmutopreco}

The search for single top quark production in the electron and muon channels
is based on the selection described in~\cite{IsoLep:2009}.
Isolated electrons and muons with a transverse momentum
$P_T^{\ell} > 10$~GeV in the polar angle range $5^{\circ}<\theta_{\ell}<140^{\circ}$
are selected in events with a missing transverse momentum
$P_T^{\rm miss} > 12$~GeV.
In the electron channel, $39$ events are found, compared to a SM prediction
of $43.1 \pm 6.0$. In the muon channel, $14$ events are observed, 
compared to a SM prediction of $11.0 \pm 1.8$. 
To estimate a potential top contribution to this sample, a top quark candidate is
reconstructed from its decay products (lepton $\ell$, neutrino $\nu$ and $b$ quark), 
and the compatibility with single top quark production via FCNC is tested
using a multivariate discriminant method.

\begin{figure}
  \begin{center}
\includegraphics[width=0.74\textwidth]{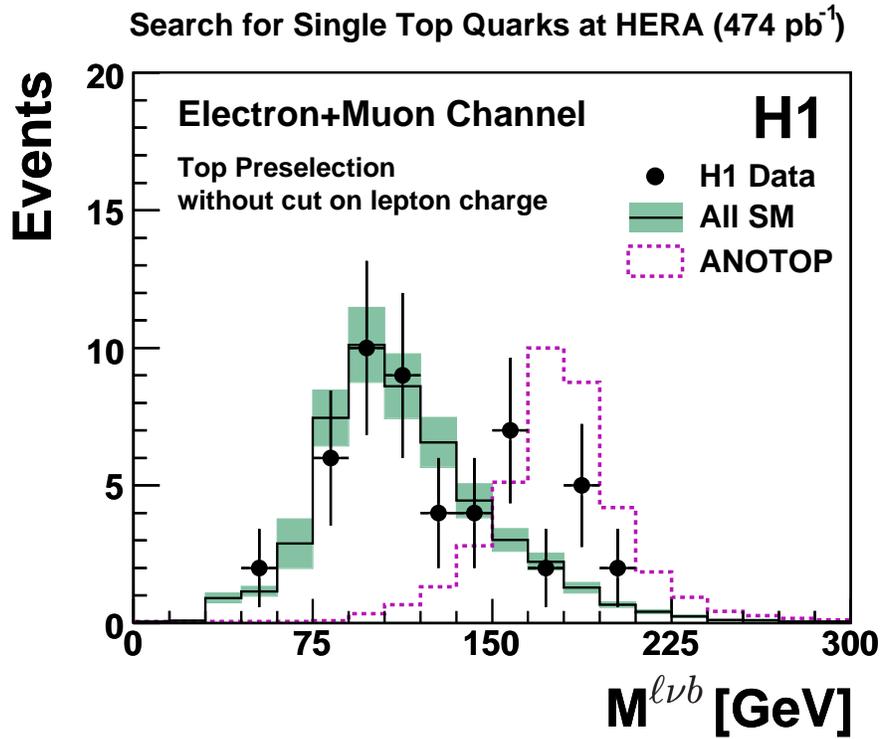}
  \end{center}
  
  \caption{Distribution of the reconstructed top mass $M^{\ell \nu b}$ 
  in the electron and muon channels after $\nu$ reconstruction but before the cut
on the lepton charge.
The data are shown as
points, the total SM expectation as the open histogram with systematic and statistical uncertainties
added in quadrature (shaded band). 
Also shown is the ANOTOP prediction with arbitrary normalisation (dashed histogram).
	}
  \label{fig:leptmtop}
\end{figure}

The neutrino four-vector $P_{\nu}$ is reconstructed using the transverse and 
longitudinal momentum balance of the event.
The transverse momentum of the neutrino is reconstructed by assuming:
\[ \vec{P}_T^{\nu} \equiv \vec{P}_T^{\rm{miss}}. \]
If the scattered electron is detected in addition to the isolated lepton,
the four-vector of the neutrino can be fully reconstructed by
exploiting the energy and longitudinal momentum balance: 
\begin{equation}
   \sum_{i} \left( E^i-P^i_z \right) + \left( E^{\nu} - P^{\nu}_z \right) 
   = 2E^{0}_e = 55.2\,\mathrm{GeV},
\label{eqn:longmom}
\end{equation}
where the sum runs over all detected particles, $P_z$ is the momentum along the proton
beam axis and $E^0_e$ is the electron beam energy.
For events with more than one isolated electron, the electron with the lower 
transverse momentum is assumed to be the scattered electron~\cite{Dingfelder2003}.
%
%
If the scattered electron is not detected, the constraint 
 $ M^{\ell\nu} \simeq M_W=\ 80.42$~GeV
is applied.
In the case that two physical solutions are obtained
for $(E^{\nu}-P^{\nu}_z)$, the more probable solution according
to the ANOTOP simulation is chosen~\cite{Dingfelder2003}.
A small fraction of events is removed in the case when the reconstruction algorithm
finds no physical solution.
In the electron channel, $38$ events are selected after this neutrino reconstruction, 
while  $39.7\pm5.6$ are expected from the SM.
In the muon channel, 13 events are selected, while $10.7\pm1.6$ are expected from the SM.
The selection efficiency for generated ANOTOP events 
at this stage is $49\%$ ($44\%$) for the electron (muon) channel.
The four-vector of the $b$-jet candidate is defined as the 
four-vector of the hadronic final state.
The four-vector of the top quark candidate is
reconstructed as the sum of the four-vectors of
the isolated lepton, the neutrino and the $b$-jet candidate.
This includes top quark decays where the $W$ boson decays
via $W \rightarrow \tau \rightarrow e (\mu)$.
Figure \ref{fig:leptmtop} shows the reconstructed top mass in the
combined electron and muon channels.
The data are in overall agreement with the SM prediction, 
while a slight excess of data events is observed in the top quark mass range.

In addition to the kinematic reconstruction of the top quark decay the lepton
charge is also exploited, as
the decay chain $t \rightarrow bW^{+} \rightarrow \ell \nu b$ produces only
positively charged leptons.
Well measured negatively charged leptons are rejected by requiring
$q_\ell\cdot\frac{|\kappa|}{\delta\kappa}>-1.0$ 
where $q_\ell$ is the charge of the lepton, $|\kappa|$ is the curvature of the track
associated to the lepton and $\delta\kappa$ is the error on the curvature.
This requirement is only applied in the central region
where the charge determination by the tracking detectors is reliable.
Table~\ref{tab:cutflow} summarises the event yield in the electron and muon channels
for the resulting top preselection.
In the electron channel $30$ events are selected, while  $31.5\pm4.0$ are 
expected from the SM.
In the muon channel, eight events are selected, while $8.0\pm0.9$ are expected from the SM.

\begin{figure}
  \begin{center}
\textsf{\large{\textbf{\quad Search for Single Top Quark Production at HERA (474 pb$^{\mbox{-1}}$) }}}

\includegraphics[width=0.49\textwidth]{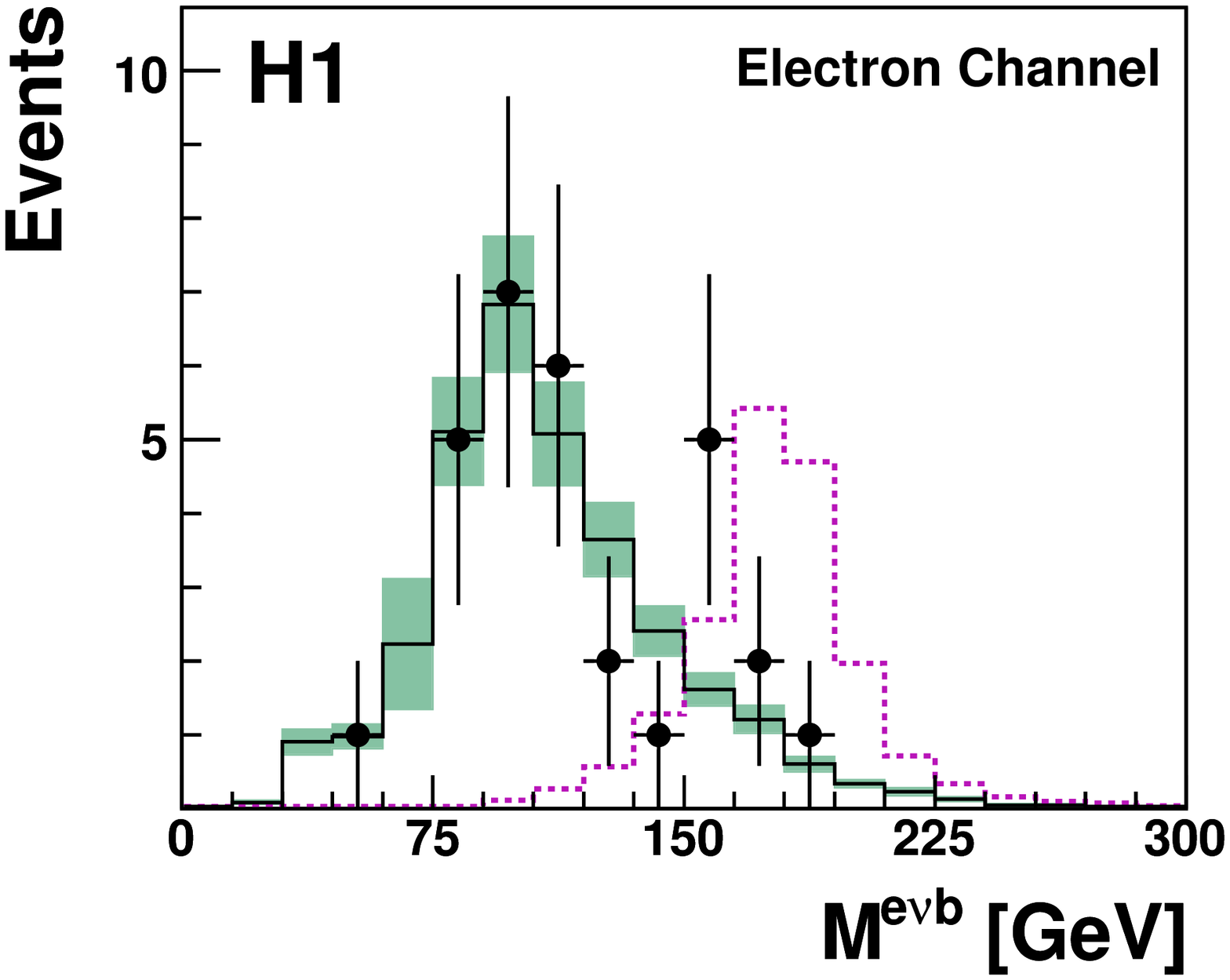}
\includegraphics[width=0.49\textwidth]{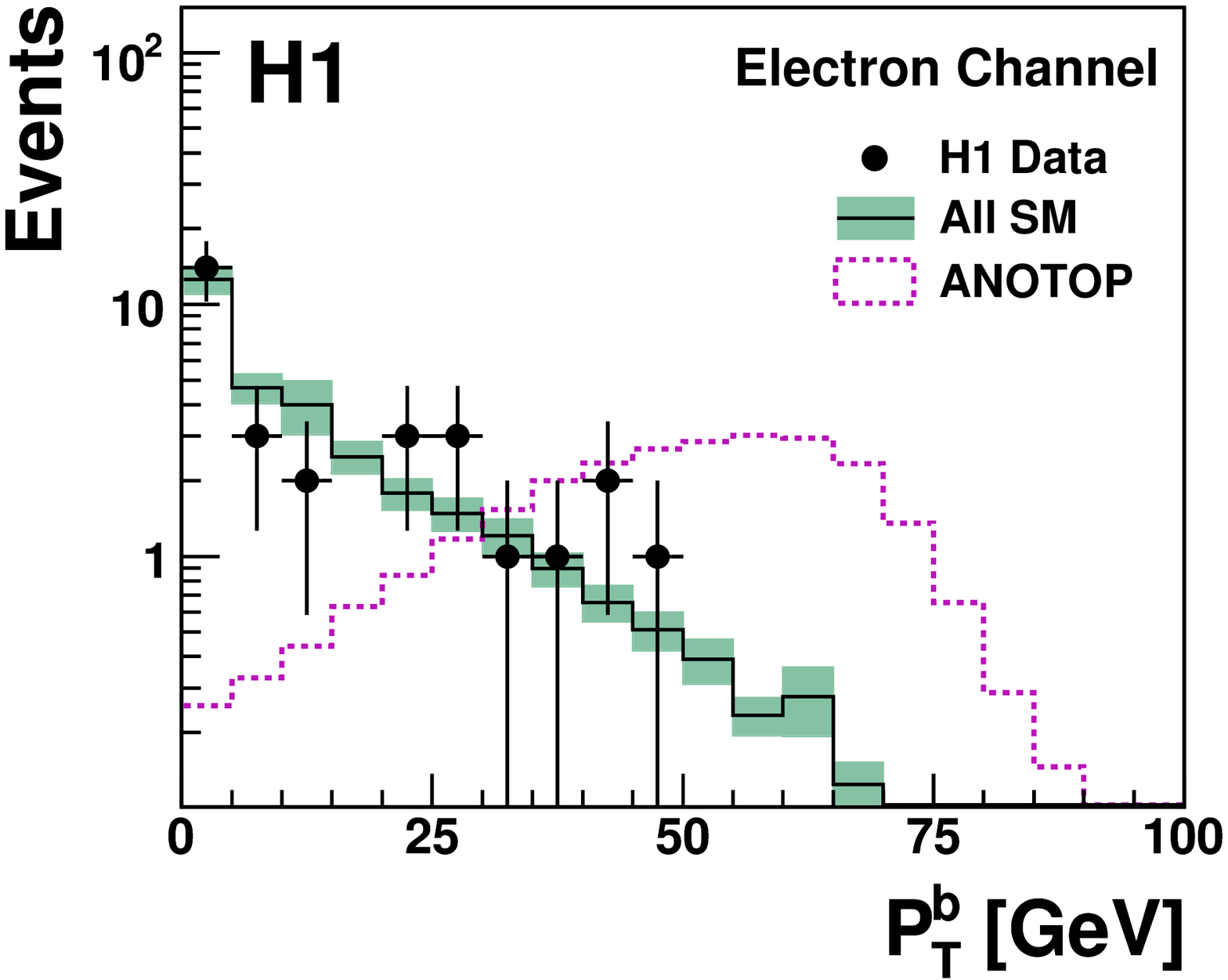}

\includegraphics[width=0.49\textwidth]{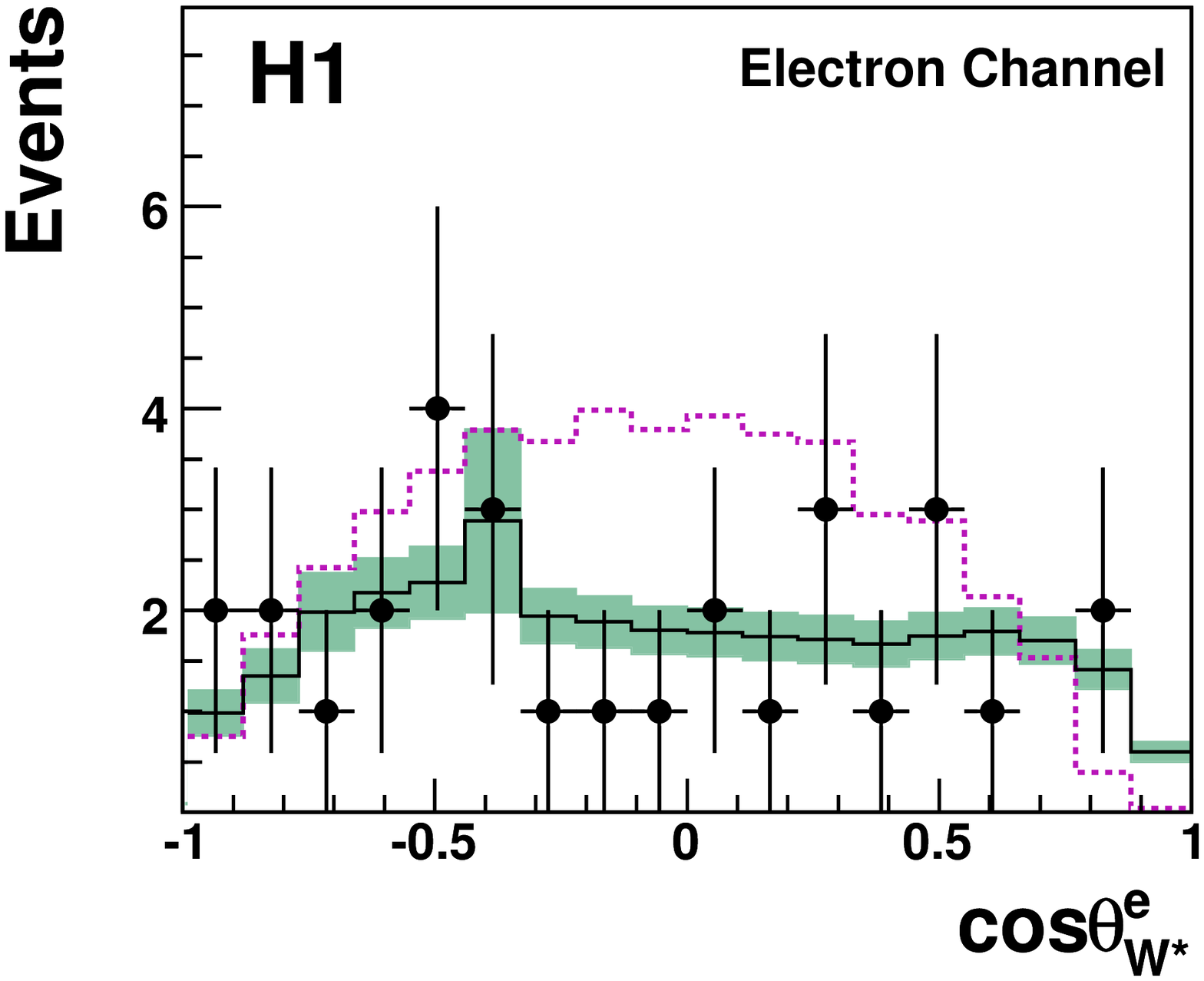}      
\psfrag{D}{}\psfrag{ \(MLP\)}{\ \ \ \ \ \ \large\it D}
\includegraphics[width=0.49\textwidth]{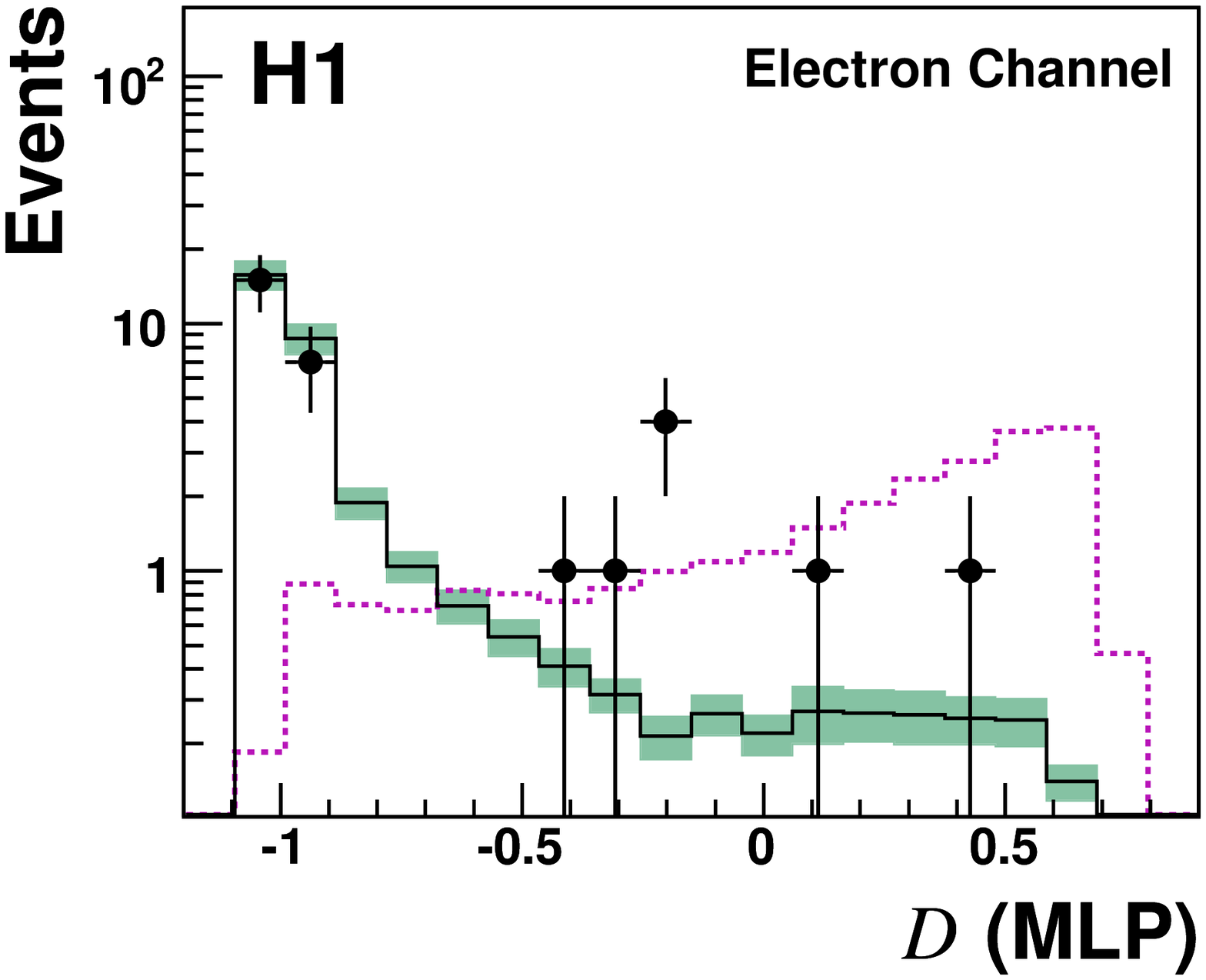}
  \end{center}
  \begin{picture} (0.,0.)
     \setlength{\unitlength}{1.0cm}
     \put ( 6.85,10.5){(a)} 
     \put (14.8 ,10.5){(b)} 
     \put ( 6.85, 5.){(c)} 
     \put (14.8 , 5.){(d)} 
  \end{picture}  
  
  \caption{Distributions of observables in the electron channel
  used to differentiate single top production from SM background processes
  in the top preselection. 
Shown are the reconstructed top mass $M^{e \nu b}$~(a),
the transverse momentum of the reconstructed $b$-jet candidate $P_T^b$~(b),
the $W$ decay angle $\cos\theta^{e}_{W^{*}}$~(c), and the combination of these
observables into a MLP--based discriminant $D$~(d). 
The data are shown as
points, the total SM expectation as the open histogram with systematic and statistical uncertainties
added in quadrature (shaded band). 
Also shown is the signal prediction with arbitrary normalisation (dashed histogram).
	}
  \label{fig:elpretop}
\end{figure}

\begin{figure}
  \begin{center}
\textsf{\large{\textbf{\quad Search for Single Top Quark Production at HERA (474 pb$^{\mbox{-1}}$) }}}\\  
      
\includegraphics[width=0.49\textwidth]{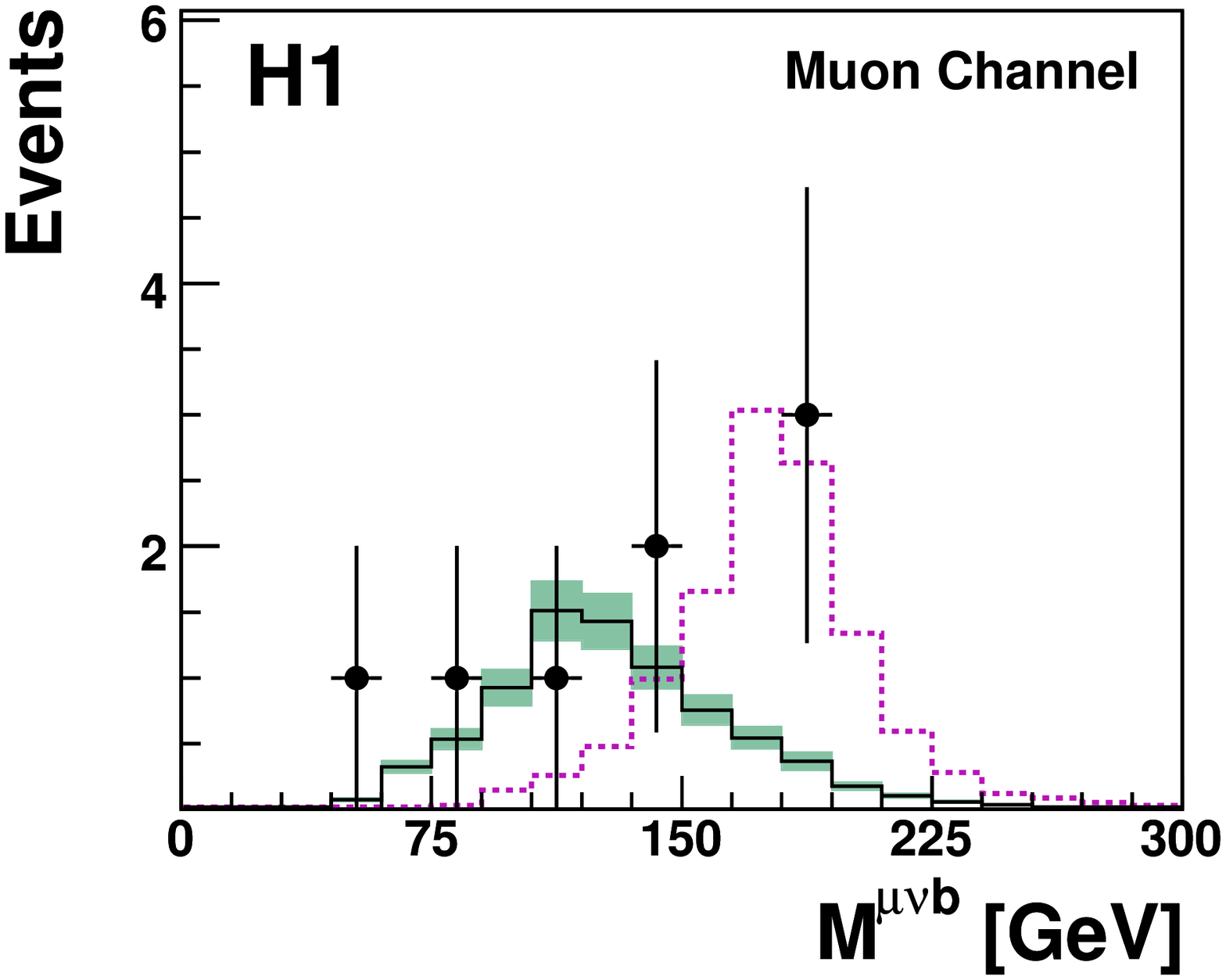}
\includegraphics[width=0.49\textwidth]{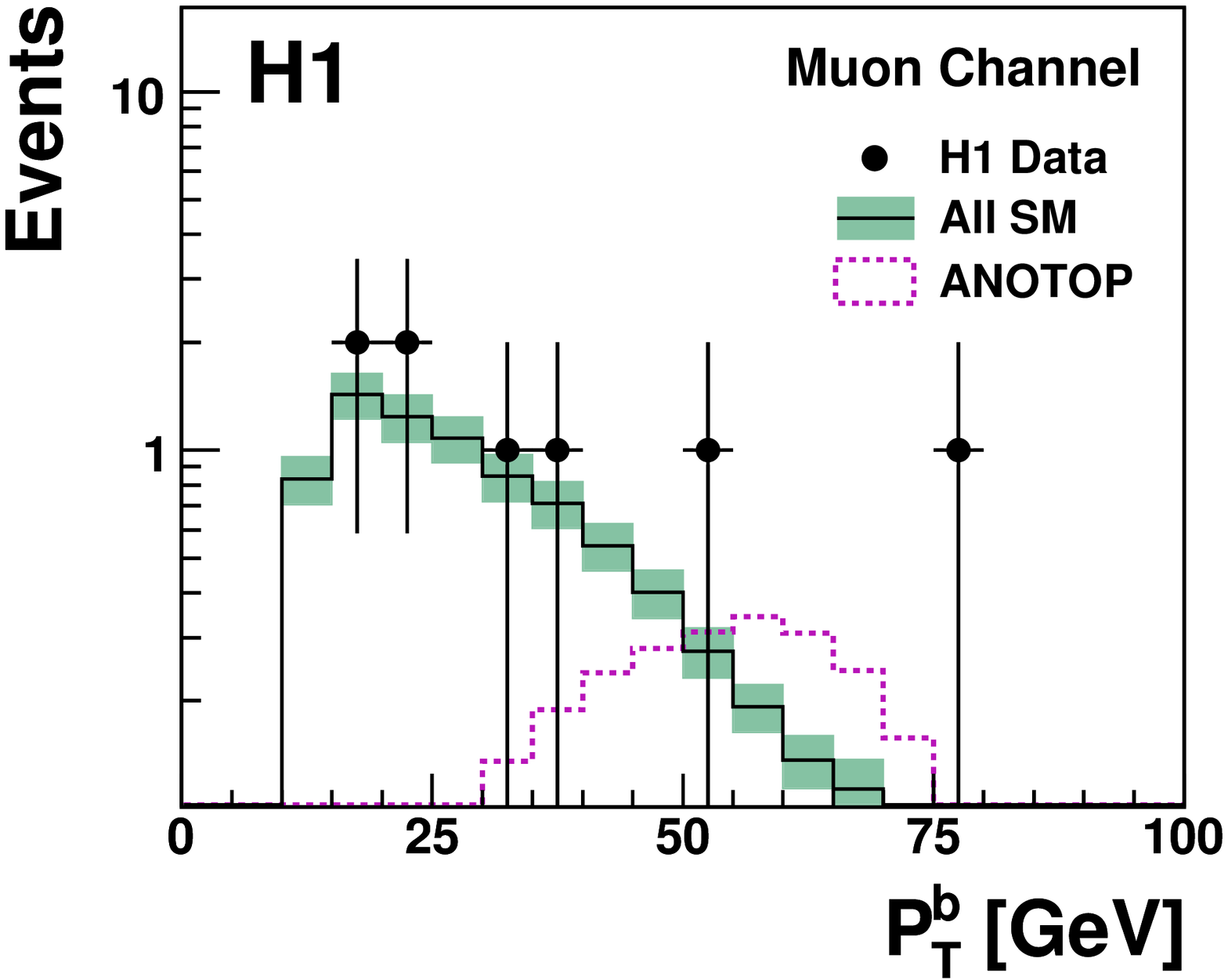}
      
\includegraphics[width=0.49\textwidth]{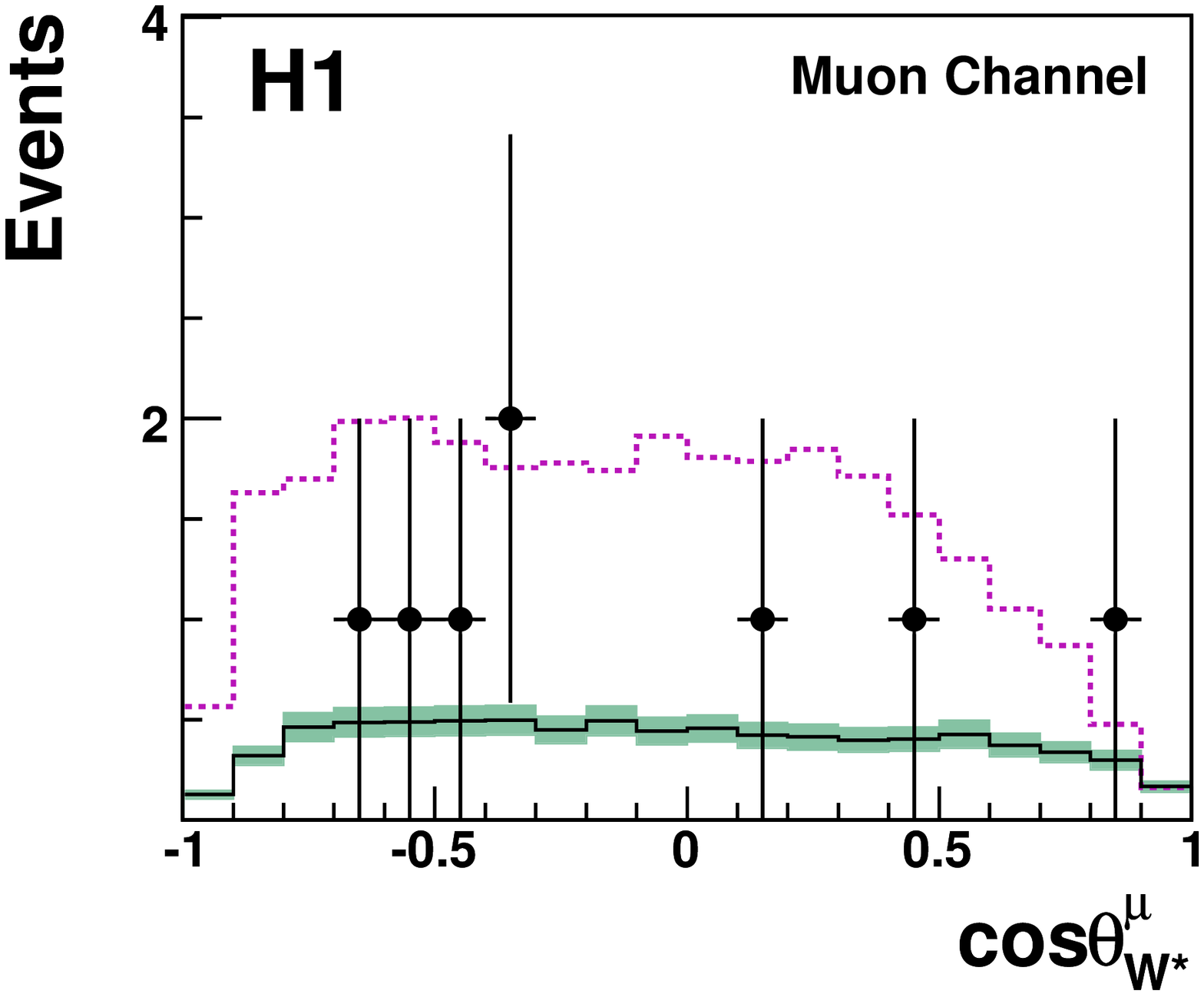}      
\psfrag{D}{}\psfrag{ \(MLP\)}{\ \ \ \ \ \ \large\it D}
\includegraphics[width=0.49\textwidth]{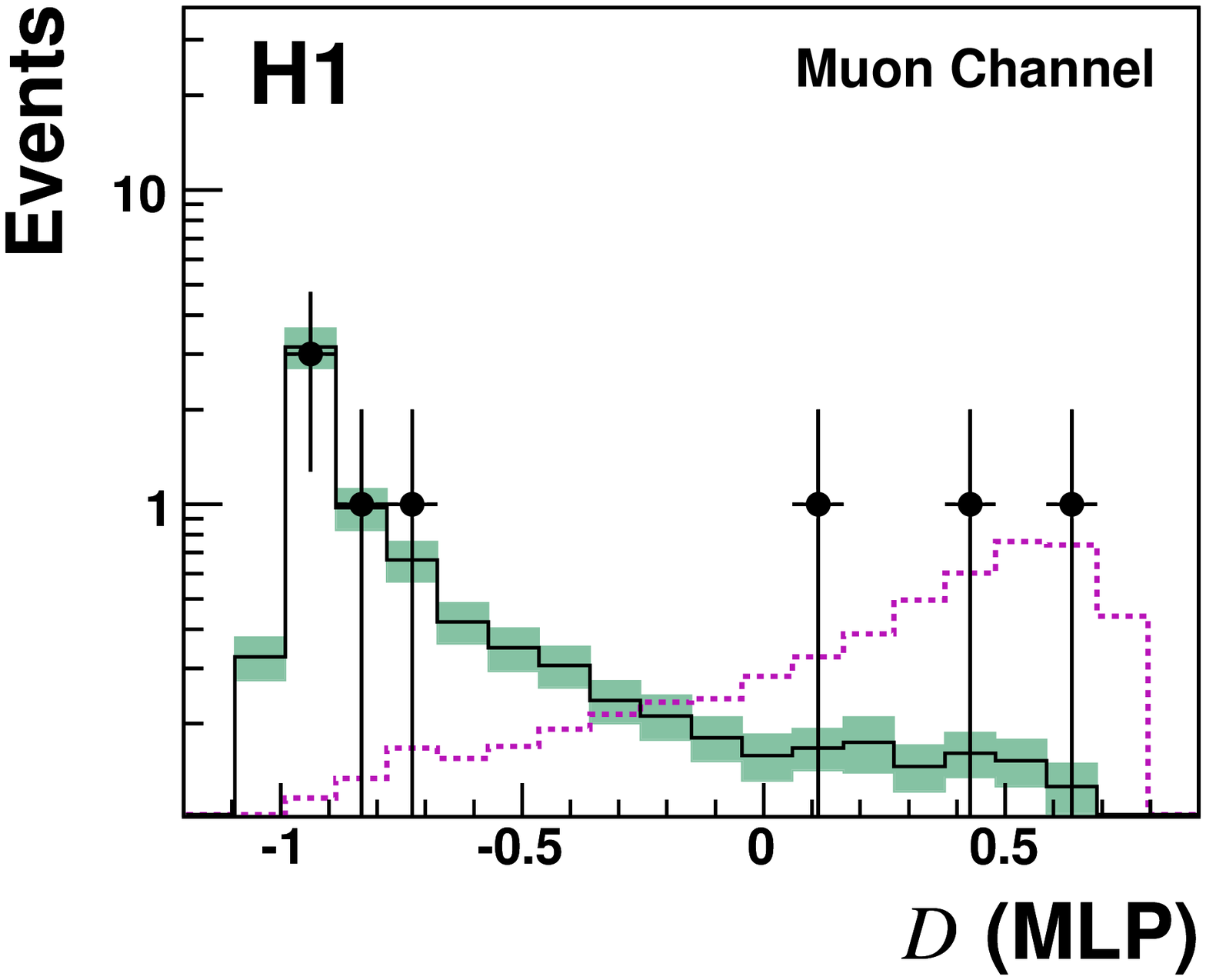}  
  \end{center}
  \begin{picture} (0.,0.)
     \setlength{\unitlength}{1.0cm}
     \put ( 6.85,11.){(a)} 
     \put (14.8 ,11.){(b)} 
     \put ( 6.85, 5.){(c)} 
     \put (14.8 , 5.){(d)} 
  \end{picture} 
  
  \caption{Distributions of observables in the muon channel 
  used to differentiate single top production from SM background processes
  in the top preselection. 
Shown are the reconstructed top mass $M^{\mu \nu b}$~(a),
the transverse momentum of the reconstructed $b$-jet candidate $P_T^b$~(b),
the $W$ decay angle $\cos\theta^{\mu}_{W^{*}}$~(c), and the combination of these
observables into a MLP--based discriminant $D$~(d). 
The data are shown as
points, the total SM expectation as the open histogram with systematic and statistical uncertainties added in quadrature (shaded band). 
Also shown is the signal prediction with arbitrary normalisation (dashed histogram).
	}
  \label{fig:mupretop}
\end{figure}

The following observables are then used
to further discriminate single top quark production via FCNC against SM background:
\begin{itemize}  
\item $P_T^b$, the transverse momentum of the $b$-jet candidate;
\item $M^{\ell \nu b}$, the invariant mass of the reconstructed top quark;
\item $\theta^{\ell}_{W^{*}}$, the $W$ decay angle calculated as the angle 
between the lepton momentum in the $W$ rest frame and the $W$ direction in 
the top quark rest frame.
\end{itemize}  
Distributions of these observables are shown in figures~\ref{fig:elpretop}~(a-c) and 
\ref{fig:mupretop}~(a-c)
for the electron and muon channels, respectively.
Good overall agreement of
the data with the SM expectation is seen in all distributions.
For the discrimination of top and background events, a multivariate discriminator based 
on a multilayer perceptron (MLP) neural network
was trained on the signal and background simulations.
Figures~\ref{fig:elpretop}~(d) and \ref{fig:mupretop}~(d) show the MLP discriminator 
output distributions for the electron and muon channels, respectively.
%
%
A good agreement is observed between data and SM expectation, with some 
events visible in the signal region.

Results from the multivariate analysis 
are cross checked with a cut-based top selection requiring for the top  
preselected events in addition \mbox{$P_T^b > 30$~GeV} and \mbox{$M^{\ell \nu b} > 140$~GeV}.
The resulting event yields are also shown in table~\ref{tab:cutflow}.
In this selection five events are selected in the electron channel, while  
$3.2 \pm 0.4$ are expected from the SM.
In the muon channel, four events are selected, while 
$2.1 \pm 0.3$ are expected from the SM.
%
%
This sample includes all five top-like events found in the previous 
analysis~\cite{Aktas:2003yd}.
If the charge requirement is not applied, the corresponding event yields
are seven selected events for an expectation of $4.1\pm0.7$ in the electron channel,
and six events for $2.8 \pm 0.4$ expected in the muon channel.

\subsection{Hadronic Channel}
\label{sec:jetstopreco}

\begin{figure}
  \begin{center}  
\textsf{\large{\textbf{\quad Search for Single Top Quark Production at HERA (474 pb$^{\mbox{-1}}$) }}}
      
\includegraphics[width=0.49\textwidth]{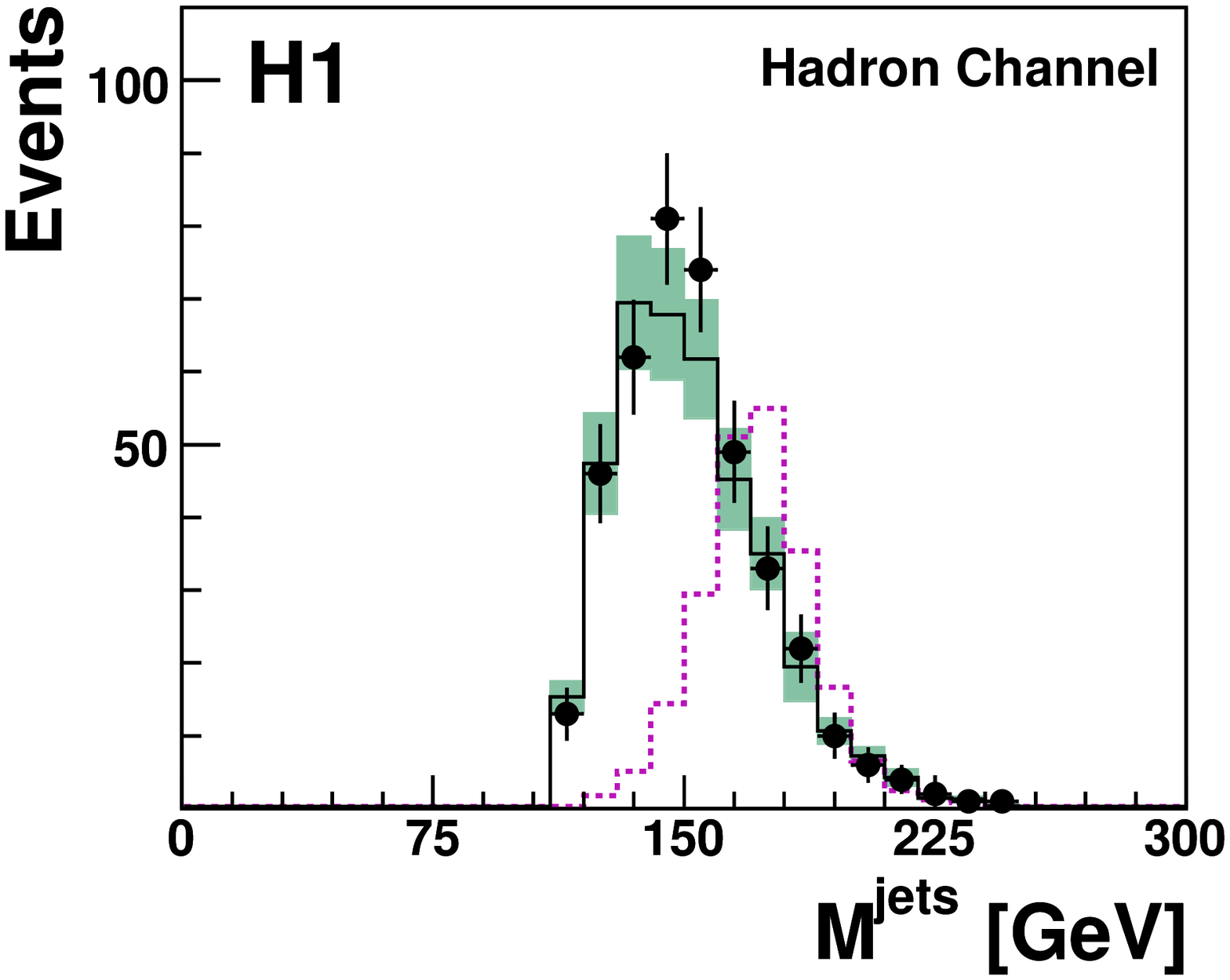}
\includegraphics[width=0.49\textwidth]{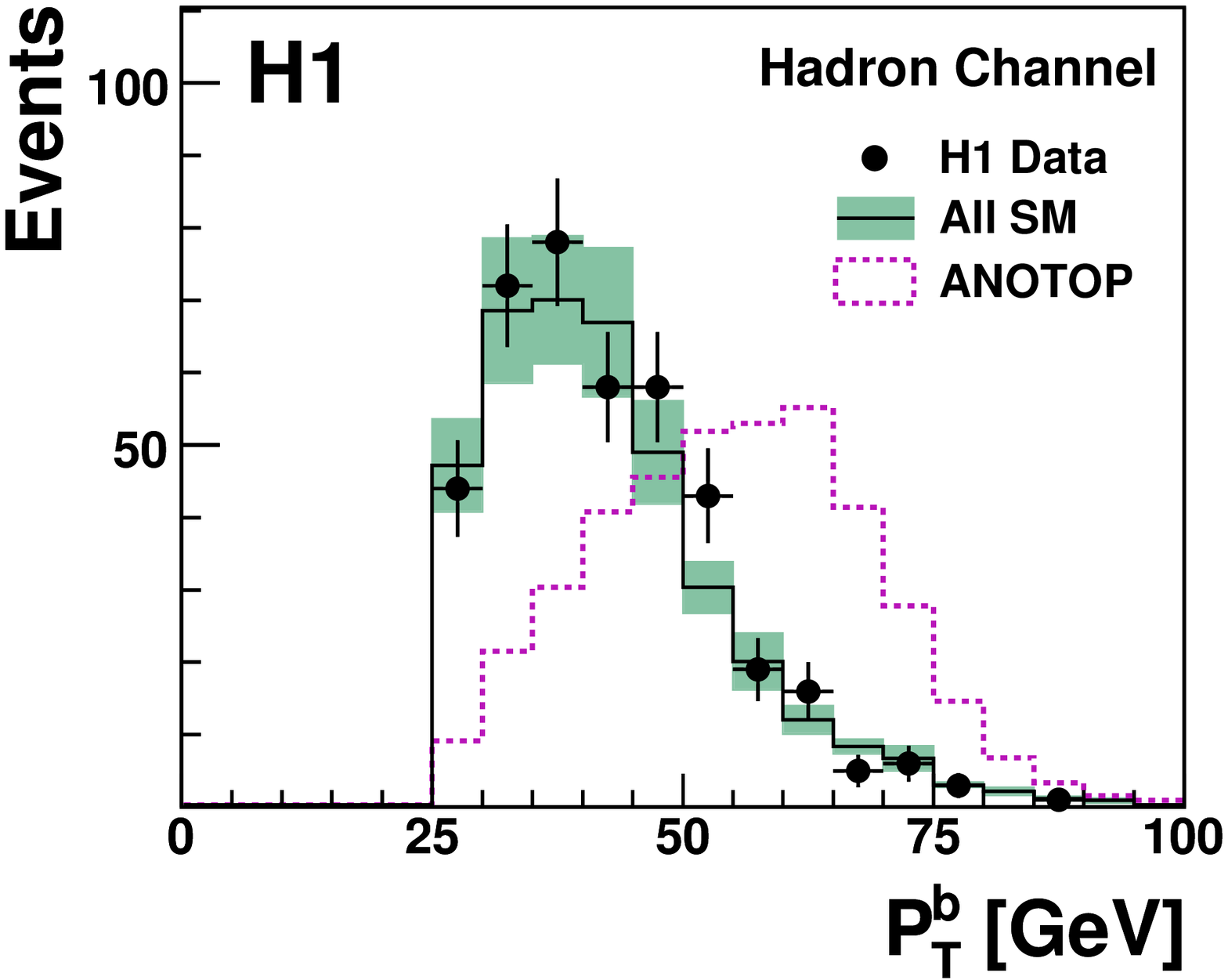}
      
\includegraphics[width=0.49\textwidth]{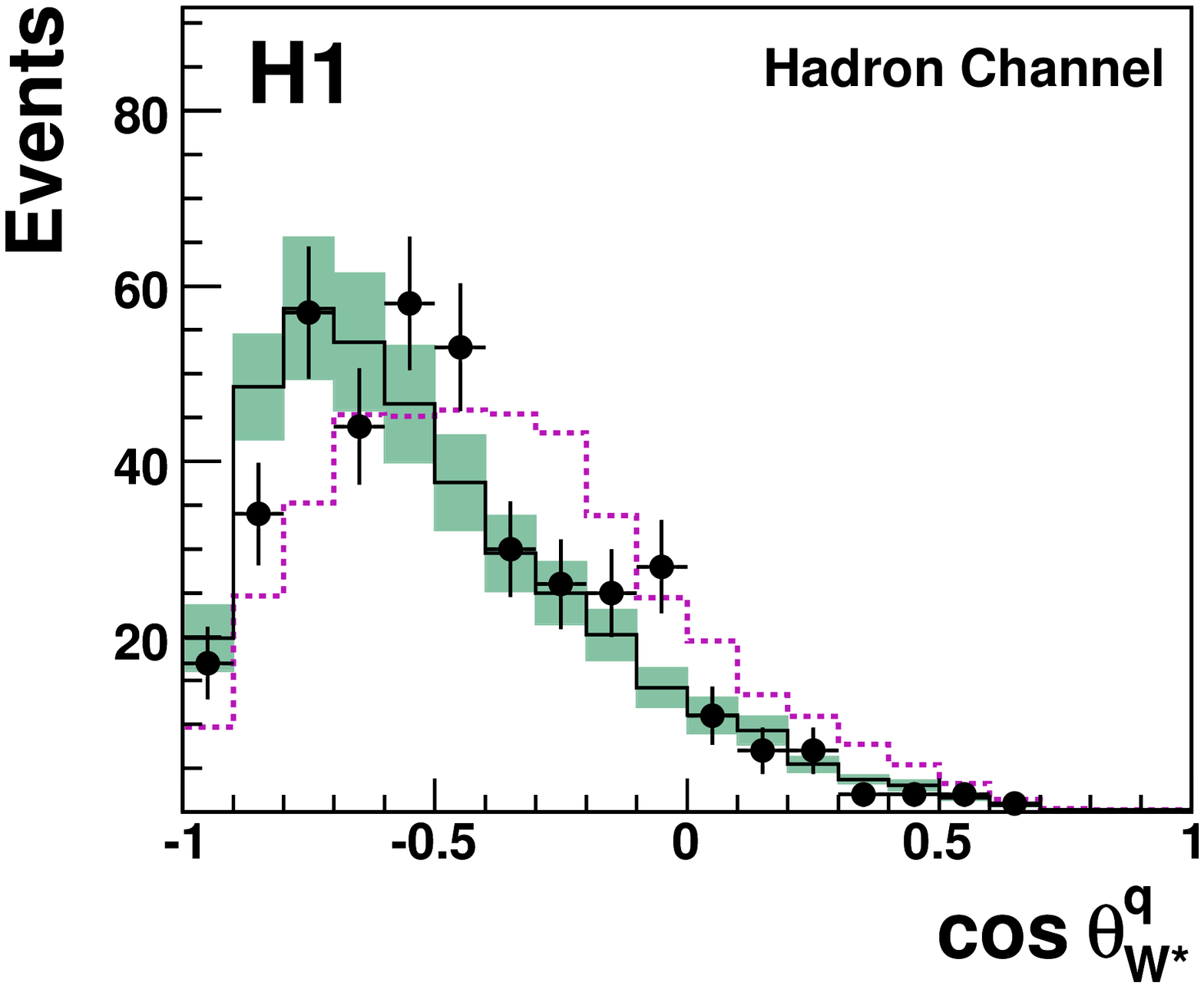}  
\psfrag{D}{}\psfrag{ \(MLP\)}{\ \ \ \ \ \ \large\it D}
\includegraphics[width=0.49\textwidth]{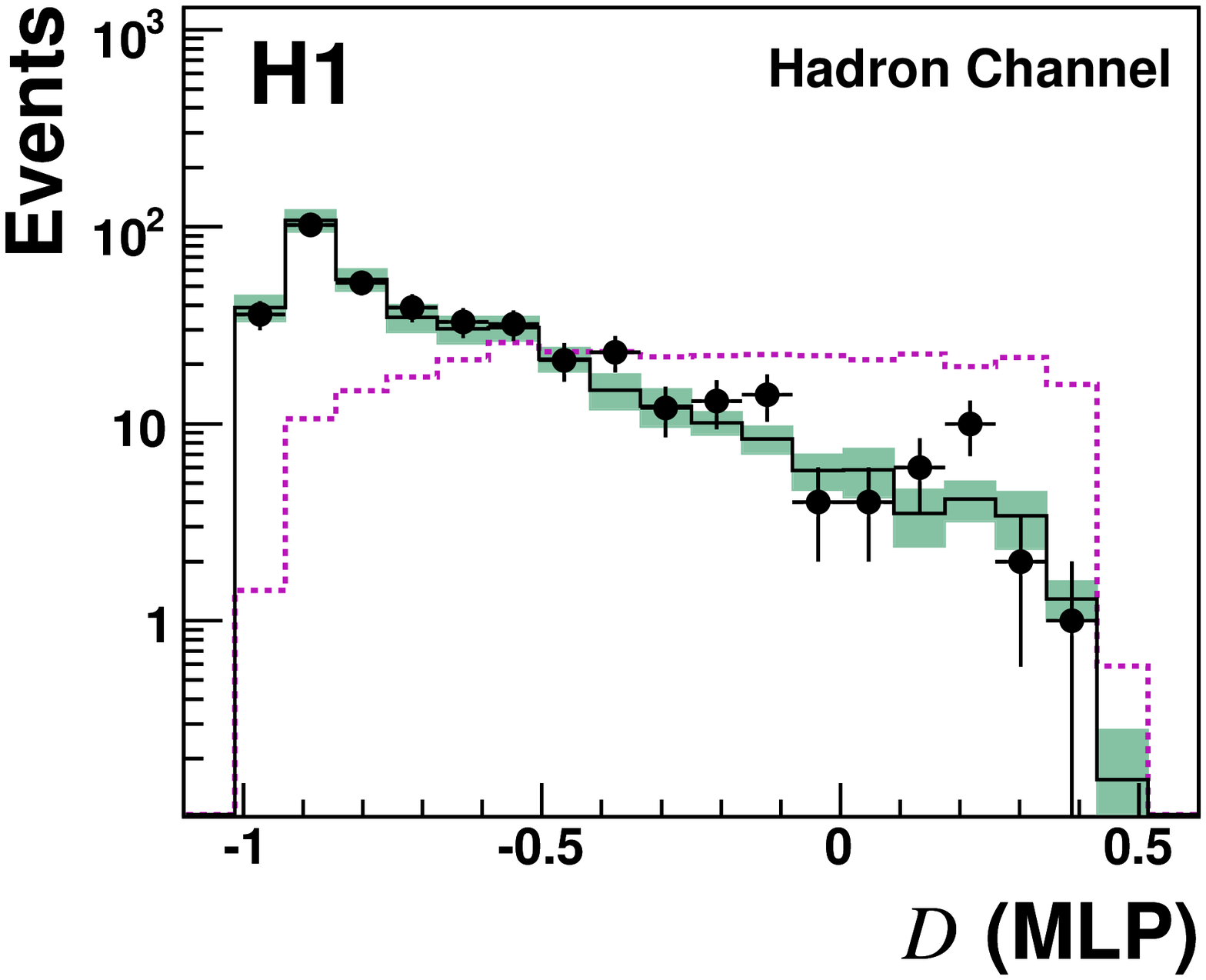}
  \end{center}    
  \begin{picture} (0.,0.)
     \setlength{\unitlength}{1.0cm}
     \put ( 6.85,10.5){(a)} 
     \put (14.8 ,10.5){(b)} 
     \put ( 6.85, 5){(c)} 
     \put (14.8 , 5){(d)} 
  \end{picture} 

  \caption{Distributions of observables in the hadronic channel 
used to differentiate single top production from SM background 
processes in the top preselection. 
Shown are the invariant mass of all jets $M^{\rm{jets}}$~(a),
the transverse momentum of the $b$-jet candidate $P_T^b$~(b),
the $W$ decay angle $\cos\theta^{q}_{W^{*}}$~(c), and the combination of these
observables into a MLP-based discriminant $D$~(d). 
The data are shown as
points, the total SM expectation as the open histogram with systematic and statistical uncertainties
added in quadrature (shaded band). 
Also shown is the signal prediction with arbitrary normalisation (dashed histogram).
}
  \label{fig:jetspretop}
\end{figure}

The hadronic decay of the $W$ boson from the top decay 
\mbox{$t\rightarrow b W \rightarrow bq\bar{q'}$} leads to
events with at least three jets with high transverse momenta. 
SM background arises mainly from multi-jet events in photoproduction or NC DIS.
A top preselection is defined by selecting
events with at least three jets
in the pseudorapidity range \mbox{$-0.5 < \eta^{\rm jet} < 2.5$}.
The jets are ordered by the magnitude of their transverse momenta and
only events with
$P_T^{\rm{jet}1}>40$~GeV, $P_T^{\rm{jet}2}>30$~GeV and $P_T^{\rm{jet}3}>15$~GeV
are selected.
A cut on the scalar sum of the jet transverse momenta $\sum P_T^{\rm jet} > 110$~GeV 
is also applied.
In addition, one of the jet pairings must yield an invariant mass between 
$65$~GeV and $95$~GeV, 
corresponding to a window around the nominal $W$ mass with a width of twice the 
mass resolution obtained for hadronic $W$ decays~\cite{Dingfelder2003}.
The remaining jet
that is not used to form the $W$ candidate is considered to originate from the $b$ quark
and is required to have a minimum $P_T>25$~GeV.
The yield for this top preselection is given in table~\ref{tab:cutflow}.
A total of $404$ events are selected, compared to 
a SM prediction of $388\pm32$ events. 
The selection efficiency for generated ANOTOP events is $42\%$ at this stage.
Analogously to the leptonic channel observables for the discrimination 
of the top quark signal from the QCD background are chosen:
\begin{itemize}
\item $P_T^{b}$, the transverse momentum of the $b$-jet candidate;
\item $M^{\rm{jets}}$, the invariant mass of all jets in the event corresponding
to the mass of the top quark for signal events;
%
\item \thstarhad, the $W$ decay angle defined as the angle in the $W$ rest frame 
between the lowest $P_T$ jet of the two jets associated to the $W$ decay and the $W$
direction in the top quark rest frame. 
\end{itemize}
Distributions of these three observables are shown in figure~\ref{fig:jetspretop}~(a-c),
compared to the SM expectation and the simulated top signal.
Good agreement between the data and the SM simulation is seen for all three 
distributions. 
For the discrimination of top and background events, an MLP discriminator 
was trained on the signal and background simulations.
Figure~\ref{fig:jetspretop}~(d) shows the discriminator output distribution for the signal
and background simulations.
Also shown are the data events as classified by the discriminator and
a good agreement is observed between data and SM expectation.
This distribution is used to extract a possible signal, as described in the 
next section.

A cut-based top selection is also performed in the hadronic channel.
The transverse momentum of the $b$-jet candidate has to fulfill  $P_T^{b} > 40$~GeV 
and the reconstructed top quark mass $150<M^{\rm{jets}}<210$~GeV. 
The number of candidate events selected is $128$,  compared with $123\pm13$
events expected from SM processes.

\section{Results}
\label{ch:results}

The discriminant distributions observed in all three channels 
are in correspondence with the SM expectation as shown in
figures~\ref{fig:elpretop}~(d), \ref{fig:mupretop}~(d) 
and \ref{fig:jetspretop}~(d).
Since no significant indication for the production of single top quarks 
is observed, upper limits on the cross section are derived 
using the method of fractional event counting which takes efficiencies, statistical 
and systematic uncertainties into account~\cite{Bock:2004xz}.
This method is cross-checked using a modified frequentist 
approach~\cite{Junk:1999kv} 
which is found to yield similar results.
Also expected cross section limits are determined 
by assuming that a top quark signal is a fluctuation of
Poisson distributed SM background. 
An ensemble of toy experiments with this
hypothesis is conducted using the background prediction 
and calculating a cross section limit at $95\%$~CL for each toy experiment.
The mean of the resulting distribution is reported as the expected cross section limit
at $95\%$~CL.

For all channels combined an upper bound on the cross section for 
single top quark production via FCNC is obtained at the $95\%$~CL:
\[ \sigma (ep \rightarrow e t X, \sqrt{s} = 319\ \rm{GeV})  < 0.25\, \mbox{pb.} \]
This cross section limit is reported for $\sqrt{s} = 319\ \rm{GeV}$, taking into account
the ratio $0.70$ of the predicted signal cross sections at $\sqrt{s} = 301\ \rm{GeV}$ and 
$319\ \rm{GeV}$~\cite{Belyaev:2001hf}.
Table~\ref{tab:toplimits} lists the observed and expected cross section limits for the
individual channels, the combined leptonic channels, and all channels combined.

The top-like events 
observed in the data at high values of the discriminant lead to
observed limits that are weaker than the expected limits. 
This difference corresponds to the slightly higher observed event yields 
from the cut-based top selection compared to the predictions.
Extracting a production cross section 
from the discriminant distributions gives for all channels combined 
$ \sigma (ep \rightarrow e t X, \sqrt{s} = 319\ \rm{GeV})  = 0.11\pm0.07$~pb. 
This value is compatible with zero within two standard deviations.

\begin{figure}[!]  
  \begin{center}
      \includegraphics[width=16cm]{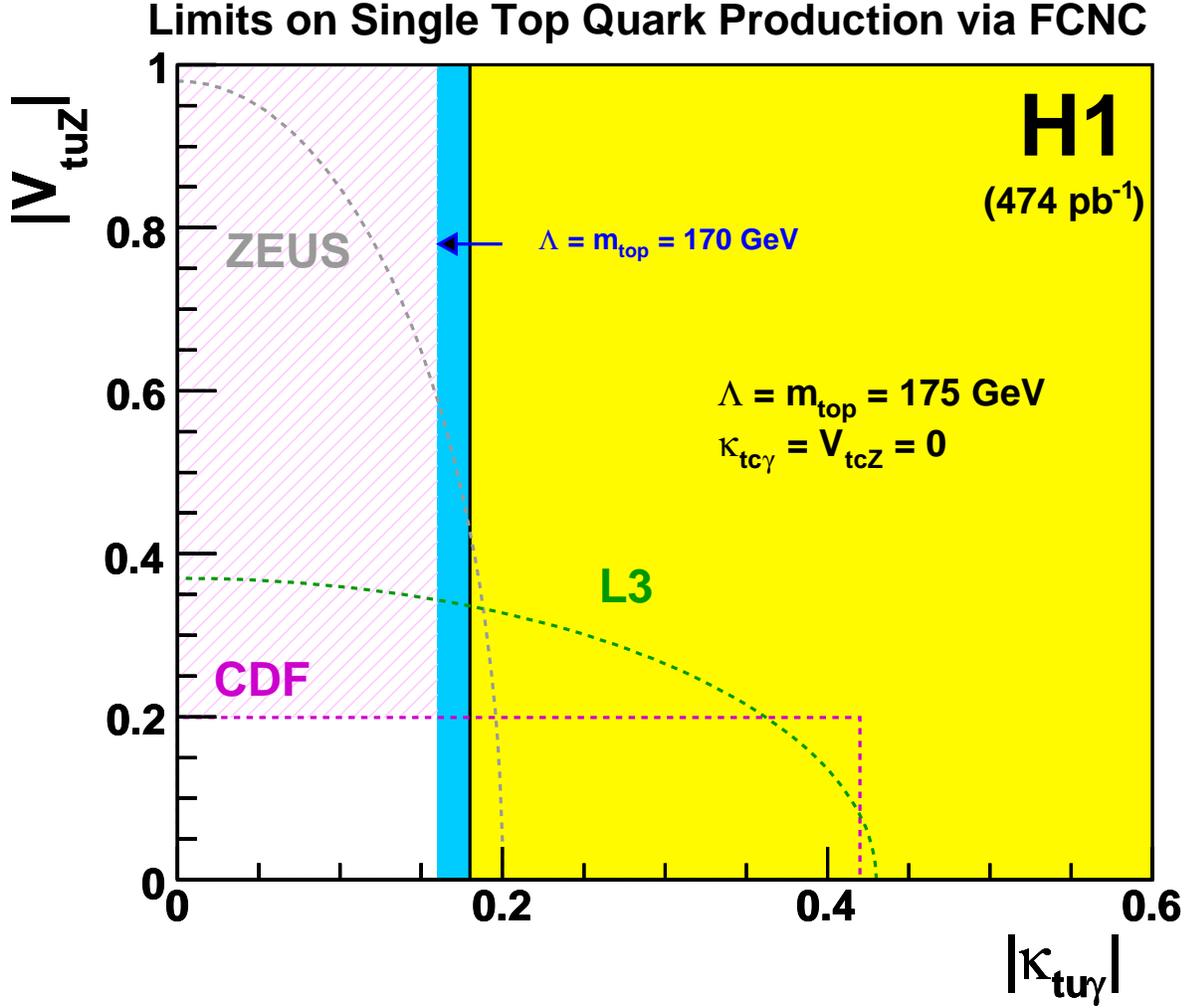}
  \end{center}
   
  \caption{Upper limits at 95\%~CL on 
the anomalous $\kappa_{tu\gamma}$ and $V_{tuZ}$ couplings obtained
assuming $\Lambda \equiv m_{\rm{top}} \equiv 175$~GeV. 
Anomalous couplings of the charm quark are neglected: \mbox{$\kappa_{tc\gamma}=V_{tcZ}=0$}.
The domain excluded by H1 is represented by the light shaded area.
The dark shaded band shows the region additionally excluded if
$\Lambda \equiv m_{\rm top}$ is varied to $170$~GeV.
The hatched area corresponds to the domain excluded at the Tevatron
by the CDF experiment~\cite{Abe:1997fz,Aaltonen:2008aaa}.
Also shown are limits obtained at LEP by the L3 experiment~\cite{Achard:2002vv}
and at HERA by the ZEUS experiment~\cite{Chekanov:2003yt}.
}
  \label{fig:top_coupling_limits}
\end{figure}

The limits on the cross section are converted to limits at $95\%$~CL on the 
anomalous FCNC coupling \ktug using equation~\ref{anotop_xsec}.
This results in an upper limit on \ktug$ < 0.16 - 0.18$ for the parameter range
$\Lambda \equiv m_{\rm top} \equiv 170-175$~GeV, respectively.
Figure~\ref{fig:top_coupling_limits} shows current limits
from experiments in the \ktug-\vtuz plane.
The limits on branching fractions 
of anomalous decays of top quarks
obtained by the CDF experiment at Tevatron~\cite{Abe:1997fz,Aaltonen:2008aaa}
are converted to limits on couplings 
using the conventions described in section 2.
%
The limit on the branching fraction $\mathcal{B} ( t \rightarrow q Z ) < 3.7\%$~at~$95\%$~CL
reported by CDF~\cite{Aaltonen:2008aaa}
corresponds to the strictest limit on the \vtuz coupling.
For ease of comparison,
the H1 limit on the coupling \ktug is also converted to a limit on the
branching fraction  $\mathcal{B}(t \rightarrow u \gamma ) < 0.64\%$,
also shown in table~\ref{tab:toplimits}.
The LEP  experiments have also searched for anomalous
single top quark production~\cite{Achard:2002vv}.
The limit obtained by the ZEUS experiment~\cite{Chekanov:2003yt} using
$130$~pb$^{-1}$ of data is similar to the H1 bound based on the full HERA data,
mainly reflecting the slight
excess of top-like events observed by H1, but not by ZEUS.
The present analysis explores a domain not covered by other colliders.

\section{Summary}
\label{ch:summary}
%
A search for single top quark production via FCNC at HERA is performed
using the full $e^{\pm}p$ data sample, corresponding to an integrated
luminosity of $474$~pb$^{-1}$.
The search is based on a sample of events with isolated leptons and 
missing transverse momentum and a sample of multi-jet events.
No evidence for single top quark production is observed.
Within FCNC models 
the top quark production cross section is extracted from multivariate
discriminator distributions using the method of fractional event counting.
An upper limit at $95\%$~CL on the production cross section:
\[ \sigma (ep \rightarrow e t X, \sqrt{s} = 319\ \rm{GeV})  < 0.25\,\rm{pb}\] 
is established. 
This result is translated into a limit on the anomalous coupling \ktug
at $95\%$~CL:
\[ \kappa_{tu\gamma} < 0.18 \]
if the scale of the new physics $\Lambda \equiv m_{\rm{top}} \equiv 175$~GeV.
This corresponds to a limit on the branching ratio
$\mathcal{B}(t \rightarrow u \gamma ) < 0.64 \%$.

\section*{Acknowledgements}

We are grateful to the HERA machine group whose outstanding efforts
have made this experiment possible. We thank the engineers and
technicians for their work in constructing and maintaining the H1
detector, our funding agencies for financial support, the DESY
technical staff for continual assistance and the DESY directorate for
support and for the hospitality which they extend to the non-DESY
members of the collaboration.

\clearpage


\begin{table} 
  \renewcommand{\arraystretch}{1.5}    
\begin{center}   
\begin{tabular}{|ccc|c|c|c|}
\hline 
\multicolumn{6}{|c|}{\bf{H1 Search for Single Top Production at HERA (\boldmath{$e^{\pm}p, 474$}~pb$^{-1}$) }} \\ 
\hline
\hline  
\multicolumn{3}{|c|}{\bf Electron Channel} & Data  & Standard Model  &  
Top Efficiency
\\
\hline  
\multirow{3}{*}{\begin{sideways}\parbox{2.1cm}{Top\newline Preselection}\end{sideways}} 
& & Isolated Leptons        & $  39 $ & $   43.1 \pm 6.0$   & $ 49 \%$ \\
& & $\nu$ Reconstruction	  & $  38 $ & $   39.7 \pm  5.6 $ & $      49 \%$ \\
& & Cut on Lepton Charge 	  & $  30 $ & $   31.5 \pm  4.0 $ & $      49 \%$ \\
\hline
& & Cut-based Top Selection & $   5 $ & $  \ 3.2 \pm  0.4 $ & $      40 \%$ \\
\hline
\hline
\multicolumn{3}{|c|}{\bf Muon Channel}  & Data  & Standard Model  &  
Top Efficiency
\\
\hline  
\multirow{3}{*}{\begin{sideways}\parbox{2.1cm}{Top\newline Preselection}\end{sideways}} 
& & Isolated Leptons         	 & $  14 $ & $  11.0 \pm  1.8 $ & $ 44 \%$ \\
& & $\nu$ Reconstruction 	 & $  13 $ & $  10.7 \pm  1.6 $ & $ 44 \%$ \\
& & Cut on Lepton Charge    	 & $   8 $ & $   8.0 \pm  0.9 $ & $ 44 \%$ \\
\hline 
& & Cut-based Top Selection & $   4 $ & $   2.1 \pm  0.3 $ & $      36 \%$ \\
\hline
\hline
\multicolumn{3}{|c|}{\bf Hadron Channel}  & Data & Standard Model &  Top Efficiency \\
\hline
& & Top Preselection	  & $ 404 $ & $  388   \pm   32  $ & $      42 \%$ \\
\hline
& & Cut-based Top Selection  & $ 128 $ & $  123   \pm   13  $ & $      33 \%$ \\
\hline
\end{tabular}  
\end{center}
   \caption{Observed and predicted numbers of events
in the electron, muon and hadronic channels for the steps of the
top preselection and in the cut-based top selection. 
The top signal efficiency is estimated using ANOTOP.
The quoted errors take into account all statistical and 
systematic uncertainties.
   } 
\label{tab:cutflow}
\end{table}
 
\clearpage

 
\begin{table}
\renewcommand{\arraystretch}{1.5}
\begin{center}
\begin{tabular}{|c|cc|c|c|} 
\hline
\multicolumn{5}{|c|}{\bf{H1 Search for Single Top Production (\boldmath{$e^{\pm}p, 474$}~pb$^{-1}$) }} \\  
\hline
\hline
 Channel &  \multicolumn{4}{|c|}{Upper Limit at $95\%$~CL} \\
\cline{2-5} 
& \multicolumn{2}{|c|}{  $ \sigma ( ep \rightarrow t X, \sqrt{s} = 319$~GeV ) }  &
  $ \kappa_{tu\gamma}$ &
  $\mathcal{B}(t \rightarrow u \gamma )$ \\
 & Observed [pb] & Expected [pb] &   & [\%]  \\ 
\hline 
Electron 	& $   0.40 $ & $   0.24 $ &  $   0.21 -   0.23 $ &  $ 0.82 - 1.02 $ \\
Muon     	& $   0.30 $ & $   0.22 $ &  $   0.18 -   0.20 $ &  $ 0.61 - 0.76 $ \\
Electron+Muon 	& $   0.27 $ & $   0.15 $ &  $   0.17 -   0.19 $ &  $ 0.55 - 0.69 $ \\
Hadronic 	& $   0.42 $ & $   0.27 $ &  $   0.21 -   0.24 $ &  $ 0.86 - 1.07 $ \\
Combined 	& $   0.25 $ & $   0.12 $ &  $   0.16 -   0.18 $ &  $ 0.51 - 0.64 $ \\
\hline
\end{tabular}
\end{center}
\caption{Observed and expected upper limits at $95\%$~CL 
on the single top production cross section at $\sqrt s=319$~GeV
in the electron, muon, hadronic and combined channels.
Also shown are upper
limits on the anomalous coupling $\kappa_{tu\gamma}$ 
and on the branching fraction $\mathcal{B}(t \rightarrow u \gamma)$.
These limits are shown for $\Lambda \equiv m_{\rm top} \equiv 170 - 175$~GeV.
   } 
\label{tab:toplimits}
\end{table}

\clearpage



\clearpage


\clearpage


\clearpage

  
\clearpage


\clearpage

\end{document}